\documentclass[reprint,amsmath,amssymb,aps]{revtex4-2}
\usepackage[T1]{fontenc}
\usepackage{graphicx}
\usepackage{dcolumn}
\usepackage{bm}
\usepackage{bbm}
\usepackage{bbold}
\usepackage{amsmath}
\usepackage{color}
\usepackage{mathtools}%

\usepackage{hyperref}

\begin{document}

\preprint{APS/123-QED}


\title{Riemann zeros and the KKR determinant}

\author{Zongrui Pei}
\affiliation{New York University, New York, NY 10003, United States}
\email{peizongrui@gmail.com; zp2137@nyu.edu}




\date{\today}

\begin{abstract}
We transform the counting function for the Riemann zeros into a Korringa–Kohn–Rostoker (KKR) determinant, assisted by Krein's theorem. This is based on our observation that the function derived from a few methods can all be recast into two terms: one corresponds to the scattering phase, and the other is similar to structure constants related to the Green function. We also discuss the possible physical realizations. Our method provides a new physical pathway towards the solution of the Riemann hypothesis.

\end{abstract}

\maketitle



\textit{Introduction}--The Riemann zeta function $\zeta(z)$ is an extension of the harmonic series $1+1/2+1/3+1/4+\dots$. It has multiple definitions, for example,
\begin{equation}
\zeta(z) = \sum_{k=1}^{+\infty} \frac{1}{k^z} = \frac{1}{\Gamma(z)}\int_0^{+\infty} dt \frac{t^{z-1}}{e^t -1}.
\end{equation}
Here, $k$ represents positive integer numbers, and $z$ is a complex number with real and imaginary parts. In the second definition, $\Gamma(z)$ is the gamma function, defined by $\Gamma(z) =\int_0^{+\infty} t^{z-1}e^{-t}dt, ~\Re (z) >0$, which is an extension of the factorial function to complex numbers. All complex numbers $z$ that satisfy the equation $\zeta(z)=0$ are called Riemann zeros.
There is a relation between $\zeta(z)$ and $\zeta(1-z)$,
\begin{equation}
\zeta(z) = 2^z \pi^{z-1} \sin(\pi z/2) \Gamma(1-z) \zeta(1-z).
\end{equation}
This equation shows that there are infinitely many trivial zeros $z=2n ~(n \in Z)$ on the real axis. In addition to these trivial zeros, there are non-trivial zeros away from the real axis. Riemann hypothesized that all these zeros are located on a line $\Re(z) =1/2$ or $z=1/2 - i E$ ($E$ is a real number), which the famous Riemann hypothesis (RH) \cite{riemann1859ueber}. RH is considered to be the most important unsolved mathematician problem. Numerical calculations hitherto show all known non-trivial zeros follow this hypothesis. However, RH cannot be proved or disproved so far.

Mathematicians search for these non-trivial zeros since they are closely connected to the distribution of prime numbers, which is usually represented by $\pi(x)$ (the number of prime numbers smaller or equal to $x$). The limit of $\pi(x)$ approaches an analytic expression, i.e.,
\begin{equation}
\lim_{x\rightarrow +\infty}\pi(x) =\frac{x}{\ln(x)},
\end{equation}
which is the prime number theorem, proved by Hadamard \cite{hadamard1896distribution} and de la Vall\'ee Poussin \cite{de1896recherches}, independently.
Riemann demonstrated that when the roots of $\zeta(z)=0$ are known, we can accurately determine $\pi(x)$.
The close relation between prime numbers and $\zeta(z)$ can also be seen more obviously from its another definition,
\begin{equation}
\zeta(z) = \prod_{p ~\mathrm{prime}} \frac{1}{1-p^{-z}}.
\end{equation}
Here, $p$ is a prime number. Assisted by the prime numbers, the zeta function is written as the multiplication of infinite terms.



\begin{table*}[]
    \centering
    \caption{The common feature of KKR determinant and various approximations to the counting function of Riemann zeros. All of them can be written in two terms $f(\vartheta) + G(E)=0$. One is related to an angle $\vartheta$ (scattering phase or undetermined wavefunction phase) due to scatters or potentials. The other term is associated with a real number $E$ ($s=1/2 +i E$) as part of a structure constant or dispersion relation. In this table, $N$ represents the counting function of the Riemann zeros.}
    \begin{tabular}{c|c|c}
    \hline \hline
    method & $f(\vartheta)$ & $G(E)$ \\
    \hline
    Riemann-Siegel formula (exact) \cite{siegel1932quell} & $f(\vartheta) =\arg \zeta(s)/\pi$  & $G(E)=\theta (E)/\pi +1$ \\
    Polya formula \cite{polya1926bemerkung} &  $f(\vartheta) = \frac{7\pi}{8} -(N+\frac{1}{2})\pi$ & $G(E) =\frac{E}{2} \ln \frac{E}{2\pi e}$ \\
    LeClair and Mussardo \cite{leclair2024riemann} & $f(\vartheta) = \vartheta -(N+\frac{1}{2})\pi$ & $G(E) = pR=E\ln \frac{E}{2\pi e}$ \\
    Sierra and Rodr\'iguez-Laguna \cite{sierra2011h} & $f(\vartheta) =-\vartheta-2\pi(N+\frac{1}{2})$ & $G(E)=\frac{E}{\hbar} \ln \frac{E}{he}$ \\
    KKR determinant of KPM \cite{Faulkner-Stocks-Yang-book} (Lloyd's formula \cite{lloyd1967wave}) & $f(\vartheta) =\cot \vartheta$ & $G(E) =\frac{\sin(\alpha a)}{\cos(ka)-\cos(\alpha a)}, \alpha=\sqrt{E}$ \\
    KKR determinant of a 1D model system (this study) & $f(\vartheta)=-t'^{-1} =-e^{-i\vartheta}$ & $ G'(E)= \Gamma(1/2+i\hat{E})/\Gamma(1/2-i\hat{E}), ~\hat{E}=E/\hbar \omega$ \\
    \hline \hline
    \end{tabular}
    \label{tab:common-feature}
    KPM: Kronig-Penny model
\end{table*}

Mathematics is an indispensable tool to develop theoretical frameworks for physics. On the other hand, Physics can also provide novel insights into the possible solutions to mathematical problems. Noteworthy examples include the solution of Poincar\'e Conjecture by Perelman \cite{perelman2002entropy} and Parisi's solution to the mean-field Sherrington-Kirkpatrick (SK) model \cite{parisi1979infinite}. Various connections between RH and physics have been proposed. Examples include (i) finding a Hamiltonian whose eigenvalues are the Riemann zeros, (ii) encoding the zero-counting function into the phase shift in a scattering process by a potential, or (iii) constructing the amplitude of scattering based on the Riemann zeta function \cite{remmen2021amplitudes}. Other proposals connect quantum chaos, quasi-crystals, the partition function of physical systems, or even the Yang-Lee zeros with zeta functions. The recent efforts have been summarized in Refs. \cite{schumayer2011colloquium,wolf2020will}. With the rise of quantum computing, a few studies have proposed to use quantum algorithms to solve RH \cite{he2020identifying,southier2023identifying,mcguigan2023quantum}.


\textit{Common feature of approximations for $N(E)$}--
The Riemann-Siegel formula for the counting function of Riemann zeros up to $E$ ($s=1/2 +i E$) is
\begin{equation}
N(E) = 1 + \frac{\theta (E)}{\pi} + \frac{1}{\pi}\arg \zeta(1/2+iE),
\end{equation}
where
\begin{equation}
\theta(t) = \frac{t}{2}\ln \bigg{(}\frac{t}{2\pi} \bigg{)} -\frac{t}{2} -\frac{\pi}{8} +\frac{1}{48t} +\frac{7}{5760t^3} +...
\end{equation}
The zero-counting function $N(E)$ can be recast into a two-term form
\begin{equation}
\label{eq:general-form}
f(\vartheta) + G(E) =0,
\end{equation}
where 
\begin{equation}
f(\vartheta) =\frac{1}{\pi}\arg \zeta(1/2+iE), ~G(E)=\frac{\theta (E)}{\pi} +1.
\end{equation}
Here, the angle $\vartheta$ represents a scattering phase or undetermined phase in wavefunctions due to potentials or scatters.
The exact $N(E)$ can be divided into two terms,
\begin{equation}
N(E)=\langle N(E)\rangle + S(E),
\end{equation}
where the first term $\langle N(E)\rangle$ is the smoothened zero-counting function, which is an approximation of $N(E)$. The second term $S(E)$ corrects the smooth function locally by the angles of $\arg \zeta(s)$. Without confusion, we still use $N(E)$ to represent $\langle N(E)\rangle$ in a few approximations introduced below. 

The common feature of a few approximations to $N(E)$ is that they can be divided into two terms. The term $f(\vartheta)$ can be treated as a scattering factor related to a potential, and the term $G(E)$ is closely related to physical quantities structure constants, dispersion relation or Green function.
An approximation to these terms based on Polya's fake $\zeta$ function \cite{polya1926bemerkung} is
\begin{equation}
\label{eq:polya}
f(\vartheta) = \frac{7\pi}{8} -(N+\frac{1}{2})\pi, ~G(E) =\frac{E}{2} \ln \frac{E}{2\pi e}.
\end{equation}
Recently, LeClair and Mussardo provided similar expressions \cite{leclair2024riemann},
\begin{equation}
f(\vartheta) = \vartheta -(N+\frac{1}{2})\pi, ~G(E) = pR=E\ln \frac{E}{2\pi e}.
\end{equation}
Here, $G(E)$ represents the dispersion relation $pR$ ($p$-momentum, $R$-lattice spacing), and the analytical expression is purely an assumption. The scattering phase can be associated with the Riemann $\zeta$ function, i.e., $\vartheta= \arg \zeta(\sigma + i E)$.
Sierra and Rodr\'iguez-Laguna proposed a revised Berry-Keating (BK) Hamiltonian \cite{sierra2011h} and found 
\begin{equation}
f(\vartheta) =-\vartheta-2\pi(N+\frac{1}{2}), ~G(E)=\frac{E}{\hbar} \ln \frac{E}{he}
\end{equation}
based on an assumed boundary condition. In the above equation, $h$ is Planck's constant, $e$ is natural constant, and $\hbar=h/2\pi$.

When the KKR theory is employed in different periodic systems, it results in determinants that also have a form of Eq. \ref{eq:general-form}. For example, in the Lloyd's formula \cite{lloyd1967wave} for the 1D Kronig-Penny model
\begin{equation}
\label{eq:1D-Kronig-Penny}
f(\vartheta) =\cot \vartheta; G(E) =\frac{\sin(\alpha a)}{\cos(ka)-\cos(\alpha a)}, \alpha=\sqrt{E}.
\end{equation}
We can find similar expressions for 2D \cite{berry1981quantizing} and 3D systems \cite{Faulkner-Stocks-Yang-book}. However, Eq. \ref{eq:1D-Kronig-Penny} and all known KKR determinants are not related to Riemann zeros and thus have expressions different from Eq. \ref{eq:polya}. In this study, we will propose a method towards the discovery of KKR determinants close to Eq. \ref{eq:polya}.


Based on the definition of $\zeta(z)$, we can find the relationship between $\zeta(z)$ and $\zeta(1-z)$ for $z=1/2 + it$,
\begin{equation}
\zeta(\frac{1}{2}-it) = \pi ^{-it} \frac{\Gamma(\frac{1}{4} + \frac{it}{2})}{\Gamma(\frac{1}{4} - \frac{it}{2})} \zeta(\frac{1}{2}+it).
\end{equation}
Any complex number can be written as $\zeta(\frac{1}{2}+it)=Z(r)e^{-i\theta(t)}$ with $Z(t)$ as a real number. Then, we can find
\begin{equation}
e^{2i\theta(t)} = \exp (-it \ln \pi) \frac{\Gamma(\frac{1}{4} + \frac{it}{2})}{\Gamma(\frac{1}{4} - \frac{it}{2})} 
\end{equation}
Then, the number of Riemann zeros can be written as
\begin{equation}
n(t) = \frac{\theta(t)}{\pi} = -\frac{t}{2\pi}\ln \pi +1 + \frac{1}{2\pi} \Im [\ln \Gamma(\frac{1}{4}+\frac{it}{2}) - \ln \Gamma(\frac{1}{4}-\frac{it}{2})].
\end{equation}
The function $n(t)$ is dominated by this term
\begin{equation}
\frac{1}{2\pi} \Im \ln \frac{\Gamma(1/4 +it/2)}{\Gamma(1/4-it/2)} \simeq \frac{t}{2\pi}\ln(\frac{t}{2e}) -\frac{1}{8} 
\end{equation}
for large $t$. We will first find a Schr\"odinger equation whose solution includes $\ln \frac{\Gamma(1/4 +it/2)}{\Gamma(1/4-it/2)}$ or a similar term $\ln \frac{\Gamma(1/2 +it)}{\Gamma(1/2-it)}$. Next, we will check the candidates of Schr\"ordinger equations and then construct a KKR determinant based on their solutions.



\begin{figure}
    \centering
    \includegraphics[width=1.0\linewidth]{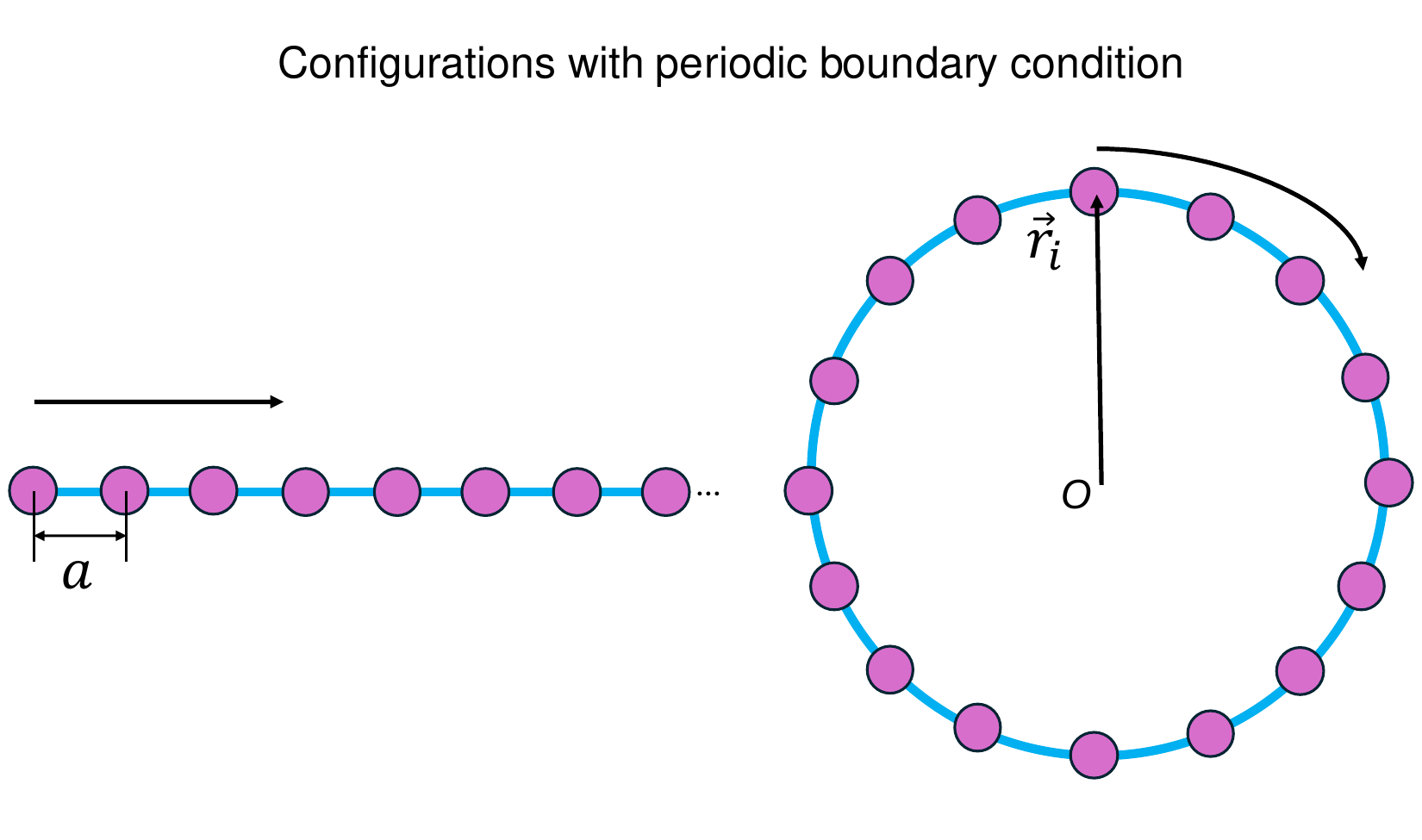}
    \caption{Schematic for two possible configurations of periodic systems. The eigenvalues of the Schr\"odinger equation correspond approximately to the Riemann zeros. The spacing $a$ between the scatters is infinitely large on the micro-scale.}
    \label{fig:schematic}
\end{figure}

There are multiple choices for a Schr\"odinger equation whose solutions embody the required term, such as the radius part of 3D Coulomb wave equations, 1D inverse Coulomb equation with a potential $-r^{-1}$ \cite{bhaduri1995phase}, or equations with a potential of the inverse harmonic oscillator $-x^2/2$ (IHO) and inverse square potential (ISP) $-r^{-2}$. ISP also appears in the anti-de Sitter/conformal field theory (AdS/CFT) theory for black hole \cite{moroz2010below}. 


The Hilbert-P\'olya conjecture states that Hamiltonians exist whose eigenvalues are Riemann zeros $z_n = 1/2 -i E_n$. Assume the Schr\"odinger equation for such a Hamiltonian $\hat{H}$ is
\begin{equation}
\hat{H} \psi_n = E_n \psi_n,
\end{equation}
where $\psi_n$ is the wavefunction with eigenvalue $E_n$.
The Schr\"odinger equation for IHO potential normalized by energy unit $\hbar \omega$ is
\begin{equation}
\label{eq:IHO}
\bigg{(}-\frac{1}{2} \frac{\partial^2}{\partial \xi^2} -\frac{1}{2} \xi^2 \bigg{)} \phi (\xi) = -\frac{E}{\hbar \omega} \phi(\xi) =-\hat{E} \phi(\xi).
\end{equation}
If we rotate the position-momentum coordinate system by $\pi/4$ and replace $-i\partial/\partial \xi +\xi \rightarrow x$ and $-i\partial/\partial \xi -\xi \rightarrow \hat{p}$, the Hamiltonian of IHO can be transformed into the following form
\begin{equation}
\hat{H} = (x\hat{p} +\hat{p} x)/2=-i\hbar \bigg{(}x\frac{d}{dx} + \frac{1}{2} \bigg{)}.
\end{equation}
Its wavefunction is of the following form
\begin{equation}
\psi_n = \alpha x^{-z_n}, ~z_n =1/2 - i E_n,
\end{equation}
where $\alpha$ is a constant. Replacing $E_n$ by $-E_n$ results in another form of wavefunction $\psi_n = \beta x^{-1/2 + iE_n}$.
This refined conjecture is also referred to as the Berry-Keating (BK) conjecture \cite{berry1999h,berry1999riemann,sierra2011h}.

We can continue to transform the squared BK Hamiltonian $[(x\hat{p}+\hat{p} x)/2]^2$ into a Schr\"odinger equation for ISP following Ref. \cite{sundaram2024duality},
\begin{equation}
\label{eq:ISP}
\bigg{(}-\frac{1}{2} \frac{\partial^2}{\partial Q^2} -\frac{g}{Q^2} \bigg{)}\psi =E\psi,
\end{equation}
where the constant $2g=\hat{E}^2 +1/4$, $Q=(-i \partial /\partial \xi +\xi)/2$ and $E=0$.
This equation can be easily solved, and its solution can be transformed into wavefunctions for the IHO or BK Hamiltonian. Following this idea, we solve Eq. \ref{eq:ISP}, transform the coordinate back to $\xi$ coordinate of the IHO model, and eventually find a solution for Eq. \ref{eq:IHO} when $\xi$ is large,
\begin{eqnarray}
\phi(\xi) \rightarrow \frac{C_1}{\sqrt{\xi}} e^{i(\frac{1}{2}\xi ^2 - \hat{E} \ln(\sqrt{2}\xi)+\frac{1}{2}\theta + \frac{\pi}{4})} \nonumber \\ + \frac{C_2}{\sqrt{\xi}} e^{-i(\frac{1}{2}\xi ^2 - \hat{E} \ln(\sqrt{2}\xi)+\frac{1}{2}\theta + \frac{\pi}{4})},
\end{eqnarray}
where the constants
\begin{equation}
C_1=\alpha e^{-\pi \frac{\hat{E}}{4}} e^{-i \frac{\pi}{8}} \Gamma(\frac{1}{2}-i\hat{E}), ~C_2=\beta e^{-\pi \frac{\hat{E}}{4}} e^{i\frac{\pi}{8}} \Gamma(\frac{1}{2}+i\hat{E}).
\end{equation}

The scattering phase factor can be calculated by the ratio of $C_1$ and $C_2$, i.e.,
\begin{equation}
e^{2i\delta_l} =\frac{C_2}{C_1} =(\beta / \alpha) \frac{\Gamma(1/2+i\hat{E})}{\Gamma(1/2-i\hat{E})} e^{i\pi/4}.
\end{equation}
Without loss of generality, the ratio of the constants can be written as another phase factor
\begin{equation}
\beta/\alpha = e^{i\vartheta}.
\end{equation}
We need a boundary condition to fix $\vartheta$. One possible way is to construct a periodic system and use the periodic boundary condition (PBC) to fix it.

{\it Krein's theorem}--
We construct a circular or linear periodic system of periodic scatters, as illustrated in Figure \ref{fig:schematic}. The Schr\"odinger equation for a circular periodic system is
\begin{equation}
\bigg{(}-\frac{\Delta}{2} -\sum_i^m\frac{g}{|\vec{r}_i|^2} \bigg{)} \psi =E\psi,  
\end{equation}
where $\Vec{r}_i= \bigg{(}r\cos(\frac{2\pi i}{m}), r\sin(\frac{2\pi i}{m})\bigg{)}$. Similarly, the Schr\"odinger equation for a 1D linear periodic system is
\begin{equation}
\bigg{(}-\frac{\Delta}{2} -\sum_i^m\frac{g}{r_i^2} \bigg{)}\psi =E\psi, 
\end{equation}
where $r_i=r-(i+1/2)a,$ and the distance between scatters $a$ is sufficiently large so that the scattering phases are independent.

Based on Krein's theorem \cite{birman1962theory,wang2014single,Galkowski2024}, we can connect a phase factor or shift $\delta_i$ with the scatter matrix $\bm{S}_i$ that associated with potential $V_i$,
\begin{equation}
\det \bm{S}_i = e^{2i \delta_i}.
\end{equation}
The total scattering phase for $m$ scatters for a 1D system (Figure \ref{fig:schematic}) is
\begin{equation}
\prod_{i=1}^m S_i =\prod_{i=1}^m e^{2i\delta_i} = (\beta/\alpha)^m \bigg{(}\frac{\Gamma(1/2+i \hat{E})}{\Gamma(1/2-i\hat{E})} \bigg{)}^m e^{i m\pi/4}.
\end{equation}
Since the wavefunction must be periodic for a periodic system, the total phase factor must satisfy this equation
\begin{equation}
\prod_{i=1}^m e^{2i\delta_i} = e^{i 2n \pi}, ~n \in Z,
\end{equation}
or equivalently,
\begin{equation}
m(\vartheta +\frac{\pi}{4}) + m \Im \ln(\frac{\Gamma(1/2+i\hat{E})}{\Gamma(1/2-i\hat{E})}) = 2n\pi.
\end{equation}

Since
\begin{equation}
\frac{1}{2\pi}\lim_{\hat{E} \rightarrow +\infty} \Im \ln(\frac{\Gamma(1/2+i\hat{E})}{\Gamma(1/2-i\hat{E})}) = \frac{\hat{E}}{\pi}\ln(\hat{E}/e),
\end{equation}
we have
\begin{equation}
\frac{n}{m} \approx \frac{1}{2\pi} (\vartheta+\frac{\pi}{4} + 2\hat{E}\ln \frac{\hat{E}}{e})
\end{equation}
when $\hat{E}$ is large. We can define $n/m$ as the counting function of the Riemann zeros $N(\hat{E})$. This equation is similar to what Sierra \textit{et al.} obtained \cite{sierra2011h}. Replacing $N(\hat{E})$ by $N(\hat{E})/2$ makes the averaged spacing of Riemann zeros closer to the exact solution.
When we compare this equation with Polya's asymptotic formula for $N(\hat{E})$ and replace $\hat{E}$ by $\hat{E}/2$, we obtain $\vartheta = 3\pi/2$. Since $\lim_{\hat{E} \rightarrow +\infty} \frac{\hat{E}/2\pi\ln(\hat{E}/2\pi e)}{\hat{E}/2\pi\ln(\hat{E}/2e)} = 1$, $N(\hat{E})$ has the same limit as Polya's formula.






{\it From $N(\hat{E})$ to KKR determinant}--
If we treat $N(\hat{E})$ as the integrated density of states for $\hat{E}$, we can use the KKR theory to connect it with a Green function $G(\hat{E})$,
\begin{equation}
\label{eq:KKR}
N(\hat{E}) =  \int_{-\infty}^{\hat{E}} n(\hat{E}')d\hat{E}', ~n(\hat{E}')= - \frac{1}{\pi} \Im G(\hat{E}').
\end{equation}
We will find an integration equation to describe $G(\hat{E})$ below.

The equation for $N(\hat{E})$ can be rewritten as
\begin{equation}
\label{eq:Krein}
N(\hat{E}) =\frac{1}{2\pi} \big{\{}\Im \ln(t') + \Im \ln(G')\big{\}},
\end{equation}
where 
\begin{equation}
t'=\beta/\alpha =e^{i\vartheta}; G'= \Gamma(1/2+i\hat{E})/\Gamma(1/2-i\hat{E}).
\end{equation}
Here, $t'$ is the phase factor. The physically meaningful scattering term usually is $t'=-\sqrt{\hat{E}} \sin \delta_l e^{i\delta_l}=i\sqrt{\hat{E}}(e^{2i\delta_l-1})/2$ for partial wave with angular momentum $l$. Then, we find $\delta_l =\arg(t') =\Im \ln(t')$. 

Eq. \ref{eq:Krein} can be transformed into
\begin{equation}
\arg(t'G') =2n \pi ~\mathrm{or}~ \Im \ln(t') + \Im \ln(G') =2n \pi.
\end{equation}
This is equivalent to
\begin{equation}
t'G'=1 ~\mathrm{or}~ t'^{-1} - G' =0.
\end{equation}
This is actually the determinant for the $1 \times 1$ matrix
\begin{equation}
\det[(t'_l)^{-1} - G'] =0.
\end{equation}
If we write this equation as $f(\vartheta)+G(\hat{E})=0$, we have $f(\vartheta)=-(t'_l)^{-1}, ~ G(\hat{E})=G'$. 
It is worth mentioning that the above equation is derived from multiple scatters, not from a single scatter. Lloyd's formula can calculate the zeros of a general KKR determinant for multiple scatters, which can be derived from Krein's theorem. It is interesting to explore this connection further in the context of the Riemann hypothesis \cite{faulkner1977scattering}.

From Eq. \ref{eq:KKR} and Eq. \ref{eq:Krein}, we find 
\begin{equation}
 \int_{-\infty}^{\hat{E}} G(\hat{E}')d\hat{E}' =-\frac{1}{2} \ln(t'G').
\end{equation}
According to KKR theory, $G(\hat{E})$ can be written as an integral of the real-space Green function $G(r,r';\hat{E})$, i.e.,
\begin{equation}
G(\hat{E})  =\int \mathrm{Tr} G(r,r';\hat{E}) drdr' = \int G(r,r;\hat{E}) dr.
\end{equation}
We will derive the explicit formula for this Green function in a separate work, which is not the primary focus of this study. Next, we will discuss the possible physical realizations of our method.


{\it Discussion}--There are multiple possibilities for the physical realizations of the scatters or potentials involved in this study.
The derivation of the key equations relies on the wavefunction to the Schr\"odinger equation with ISP, so we first discuss how to construct this potential. One possible pathway is to construct a composite scatter at each site, i.e., each scatter consists of a negative charge $-q$ at $\vec{r}$ and positive charge $+q$ separated by a distance of $|\vec{a}_0|$,
\begin{equation}
V(r)=\frac{q}{|\vec{r}|} -\frac{q}{|\Vec{r}+\Vec{a}_0|} \sim -\frac{q}{r^2}, ~\mathrm{with}~ a_0 \ll r.
\end{equation}
The choice of potentials and Schr\"odinger equations are not unique. Other discrete scatters or potentials can be used if they can provide the required scattering phase factor but keep magnitude invariant.
For example, we can construct systems with a Hamiltonian like IHO $-\frac{1}{2} m\omega^2 x^2$ or BK and then use mathematical tricks to transform them into ISP $1/r^2$. 
The BK Hamiltonian is closely related to the Dirac Hamiltonian \cite{gupta2013dirac,rodriguez2017synthetic}, such as through the Rindler coordinate transform, which provides additional physical methods (e.g., cold atoms) to validate the RH.


When we use random on-site potentials $V_i$, the system is similar to spin glass. Quantum-mechanical treatment of this problem results in a random matrix theory (RMT) for the eigenvalues. Randomness is an essential feature of the Riemann zeros, which connects RMT with RH. Spin glass could provide a physical realization to solve RH.

In addition to relying on the scattering effect of discrete scatters, the collective behavior of exotic condensed matter is promising. The condensed matter has been shown to be a rich field that helps identify particles or, more precisely, quasi-particles predicted in high-energy physics \cite{elliott2015colloquium,flensberg2021engineered}. 
The topologically protected properties come from the phase factor of wavefunctions, which is determined by the band structure of condensed matter. The pseudo-magnetism of 2D topological materials can manipulate the phase factor $S=e^{2i\delta_l}$ or $e^{i\gamma}$. Here the phase factor $\gamma(C)=q/\hbar \oint_C \bm{A}(\bm{R})=q\Phi/\hbar$ with $\bm{A}(\bm{R})$ as the vector potential in space $\bm{R}$ \cite{berry1984quantal}. In this case $N(\hat{E})$ will be connected with the Chern number $C$. This offers a platform and opportunity to study the phase factor $e^{2i\delta_l}$ in the $\zeta$ function and its zeros. KKR theory provides a physical method to find band structures as a solution to KKR determinants. So, with the help of the theory and the use of 2D quantum materials as a medium, we may solve the RH.

{\it Conclusions}--We proposed a model system to connect the KKR determinant with the counting function of Riemann zeros. We also discuss the possible physical realization of the model system. More specifically, we start with the solution of the Schr\"odinger equation for inverse square potential (ISP) and then transform its solution into the wavefuntion for $x\hat{p}$ Hamiltonian. We then construct a model Hamiltonian for ISP scatters in periodic patterns to use the periodic boundary condition to fix a phase factor. Solutions to this problem result in one equation for phase factors (assume only phase factors change), which is a function of the gamma function. The phase factor is transformed to $N(\hat{E})$. The major result of this work is that the equation for $N(\hat{E})$ can be written as a KKR determinant. Future work will find the Green function $G(\hat{E},r,r')$ and a real material with the properties. 




%





\begin{thebibliography}{32}%
\makeatletter
\providecommand \@ifxundefined [1]{%
 \@ifx{#1\undefined}
}%
\providecommand \@ifnum [1]{%
 \ifnum #1\expandafter \@firstoftwo
 \else \expandafter \@secondoftwo
 \fi
}%
\providecommand \@ifx [1]{%
 \ifx #1\expandafter \@firstoftwo
 \else \expandafter \@secondoftwo
 \fi
}%
\providecommand \natexlab [1]{#1}%
\providecommand \enquote  [1]{``#1''}%
\providecommand \bibnamefont  [1]{#1}%
\providecommand \bibfnamefont [1]{#1}%
\providecommand \citenamefont [1]{#1}%
\providecommand \href@noop [0]{\@secondoftwo}%
\providecommand \href [0]{\begingroup \@sanitize@url \@href}%
\providecommand \@href[1]{\@@startlink{#1}\@@href}%
\providecommand \@@href[1]{\endgroup#1\@@endlink}%
\providecommand \@sanitize@url [0]{\catcode `\\12\catcode `\$12\catcode `\&12\catcode `\#12\catcode `\^12\catcode `\_12\catcode `\%12\relax}%
\providecommand \@@startlink[1]{}%
\providecommand \@@endlink[0]{}%
\providecommand \url  [0]{\begingroup\@sanitize@url \@url }%
\providecommand \@url [1]{\endgroup\@href {#1}{\urlprefix }}%
\providecommand \urlprefix  [0]{URL }%
\providecommand \Eprint [0]{\href }%
\providecommand \doibase [0]{https://doi.org/}%
\providecommand \selectlanguage [0]{\@gobble}%
\providecommand \bibinfo  [0]{\@secondoftwo}%
\providecommand \bibfield  [0]{\@secondoftwo}%
\providecommand \translation [1]{[#1]}%
\providecommand \BibitemOpen [0]{}%
\providecommand \bibitemStop [0]{}%
\providecommand \bibitemNoStop [0]{.\EOS\space}%
\providecommand \EOS [0]{\spacefactor3000\relax}%
\providecommand \BibitemShut  [1]{\csname bibitem#1\endcsname}%
\let\auto@bib@innerbib\@empty
\bibitem [{\citenamefont {Riemann}(1859)}]{riemann1859ueber}%
  \BibitemOpen
  \bibfield  {author} {\bibinfo {author} {\bibfnamefont {B.}~\bibnamefont {Riemann}},\ }\bibfield  {title} {\bibinfo {title} {Ueber die anzahl der primzahlen unter einer gegebenen grosse},\ }\href@noop {} {\bibfield  {journal} {\bibinfo  {journal} {Ges. Math. Werke und Wissenschaftlicher Nachla{\ss}}\ }\textbf {\bibinfo {volume} {2}},\ \bibinfo {pages} {2} (\bibinfo {year} {1859})}\BibitemShut {NoStop}%
\bibitem [{\citenamefont {Hadamard}(1896)}]{hadamard1896distribution}%
  \BibitemOpen
  \bibfield  {author} {\bibinfo {author} {\bibfnamefont {J.}~\bibnamefont {Hadamard}},\ }\bibfield  {title} {\bibinfo {title} {Sur la distribution des z{\'e}ros de la fonction $\zeta (s) $ et ses cons{\'e}quences arithm{\'e}tiques},\ }\href@noop {} {\bibfield  {journal} {\bibinfo  {journal} {Bulletin de la Societ{\'e} mathematique de France}\ }\textbf {\bibinfo {volume} {24}},\ \bibinfo {pages} {199} (\bibinfo {year} {1896})}\BibitemShut {NoStop}%
\bibitem [{\citenamefont {De~La Vallee-Poussin}(1896)}]{de1896recherches}%
  \BibitemOpen
  \bibfield  {author} {\bibinfo {author} {\bibfnamefont {C.}~\bibnamefont {De~La Vallee-Poussin}},\ }\bibfield  {title} {\bibinfo {title} {Recherches analytiques sur la th{\'e}orie des nombres premiers},\ }\href@noop {} {\bibfield  {journal} {\bibinfo  {journal} {Ann. Soc. Sc. Bruxelles}\ } (\bibinfo {year} {1896})}\BibitemShut {NoStop}%
\bibitem [{\citenamefont {Siegel}\ and\ \citenamefont {zur~analytischen Zahlentheorie}(1932)}]{siegel1932quell}%
  \BibitemOpen
  \bibfield  {author} {\bibinfo {author} {\bibfnamefont {C.}~\bibnamefont {Siegel}}\ and\ \bibinfo {author} {\bibfnamefont {{\"U}.~R.~N.}\ \bibnamefont {zur~analytischen Zahlentheorie}},\ }\bibfield  {title} {\bibinfo {title} {{\guillemotright}, quell},\ }\href@noop {} {\bibfield  {journal} {\bibinfo  {journal} {Stud. Gesch. Math. B}\ }\textbf {\bibinfo {volume} {2}},\ \bibinfo {pages} {45} (\bibinfo {year} {1932})}\BibitemShut {NoStop}%
\bibitem [{\citenamefont {P{\'o}lya}(1926)}]{polya1926bemerkung}%
  \BibitemOpen
  \bibfield  {author} {\bibinfo {author} {\bibfnamefont {G.}~\bibnamefont {P{\'o}lya}},\ }\bibfield  {title} {\bibinfo {title} {Bemerkung {\"u}ber die integraldarstellung der riemannschen $\xi$-funktion},\ }\href@noop {} {\bibfield  {journal} {\bibinfo  {journal} {Acta Mathematica}\ }\textbf {\bibinfo {volume} {48}},\ \bibinfo {pages} {305} (\bibinfo {year} {1926})}\BibitemShut {NoStop}%
\bibitem [{\citenamefont {LeClair}\ and\ \citenamefont {Mussardo}(2024)}]{leclair2024riemann}%
  \BibitemOpen
  \bibfield  {author} {\bibinfo {author} {\bibfnamefont {A.}~\bibnamefont {LeClair}}\ and\ \bibinfo {author} {\bibfnamefont {G.}~\bibnamefont {Mussardo}},\ }\bibfield  {title} {\bibinfo {title} {Riemann zeros as quantized energies of scattering with impurities},\ }\href@noop {} {\bibfield  {journal} {\bibinfo  {journal} {Journal of High Energy Physics}\ }\textbf {\bibinfo {volume} {2024}},\ \bibinfo {pages} {1} (\bibinfo {year} {2024})}\BibitemShut {NoStop}%
\bibitem [{\citenamefont {Sierra}\ and\ \citenamefont {Rodr{\'\i}guez-Laguna}(2011)}]{sierra2011h}%
  \BibitemOpen
  \bibfield  {author} {\bibinfo {author} {\bibfnamefont {G.}~\bibnamefont {Sierra}}\ and\ \bibinfo {author} {\bibfnamefont {J.}~\bibnamefont {Rodr{\'\i}guez-Laguna}},\ }\bibfield  {title} {\bibinfo {title} {H= x p model revisited and the riemann zeros},\ }\href@noop {} {\bibfield  {journal} {\bibinfo  {journal} {Physical review letters}\ }\textbf {\bibinfo {volume} {106}},\ \bibinfo {pages} {200201} (\bibinfo {year} {2011})}\BibitemShut {NoStop}%
\bibitem [{\citenamefont {Faulkner}\ \emph {et~al.}(2018)\citenamefont {Faulkner}, \citenamefont {Stocks},\ and\ \citenamefont {Wang}}]{Faulkner-Stocks-Yang-book}%
  \BibitemOpen
  \bibfield  {author} {\bibinfo {author} {\bibfnamefont {J.~S.}\ \bibnamefont {Faulkner}}, \bibinfo {author} {\bibfnamefont {G.~M.}\ \bibnamefont {Stocks}},\ and\ \bibinfo {author} {\bibfnamefont {Y.}~\bibnamefont {Wang}},\ }\bibfield  {title} {\bibinfo {title} {Toy models},\ }in\ \href {https://doi.org/10.1088/2053-2563/aae7d8ch7} {\emph {\bibinfo {booktitle} {Multiple Scattering Theory}}},\ \bibinfo {series and number} {2053-2563}\ (\bibinfo  {publisher} {IOP Publishing},\ \bibinfo {year} {2018})\ pp.\ \bibinfo {pages} {7--1 to 7--14}\BibitemShut {NoStop}%
\bibitem [{\citenamefont {Lloyd}(1967)}]{lloyd1967wave}%
  \BibitemOpen
  \bibfield  {author} {\bibinfo {author} {\bibfnamefont {P.}~\bibnamefont {Lloyd}},\ }\bibfield  {title} {\bibinfo {title} {Wave propagation through an assembly of spheres: Ii. the density of single-particle eigenstates},\ }\href@noop {} {\bibfield  {journal} {\bibinfo  {journal} {Proceedings of the Physical Society}\ }\textbf {\bibinfo {volume} {90}},\ \bibinfo {pages} {207} (\bibinfo {year} {1967})}\BibitemShut {NoStop}%
\bibitem [{\citenamefont {Perelman}(2002)}]{perelman2002entropy}%
  \BibitemOpen
  \bibfield  {author} {\bibinfo {author} {\bibfnamefont {G.}~\bibnamefont {Perelman}},\ }\bibfield  {title} {\bibinfo {title} {The entropy formula for the ricci flow and its geometric applications},\ }\href@noop {} {\bibfield  {journal} {\bibinfo  {journal} {arXiv preprint math/0211159}\ } (\bibinfo {year} {2002})}\BibitemShut {NoStop}%
\bibitem [{\citenamefont {Parisi}(1979)}]{parisi1979infinite}%
  \BibitemOpen
  \bibfield  {author} {\bibinfo {author} {\bibfnamefont {G.}~\bibnamefont {Parisi}},\ }\bibfield  {title} {\bibinfo {title} {Infinite number of order parameters for spin-glasses},\ }\href@noop {} {\bibfield  {journal} {\bibinfo  {journal} {Physical Review Letters}\ }\textbf {\bibinfo {volume} {43}},\ \bibinfo {pages} {1754} (\bibinfo {year} {1979})}\BibitemShut {NoStop}%
\bibitem [{\citenamefont {Remmen}(2021)}]{remmen2021amplitudes}%
  \BibitemOpen
  \bibfield  {author} {\bibinfo {author} {\bibfnamefont {G.~N.}\ \bibnamefont {Remmen}},\ }\bibfield  {title} {\bibinfo {title} {Amplitudes and the riemann zeta function},\ }\href@noop {} {\bibfield  {journal} {\bibinfo  {journal} {Physical Review Letters}\ }\textbf {\bibinfo {volume} {127}},\ \bibinfo {pages} {241602} (\bibinfo {year} {2021})}\BibitemShut {NoStop}%
\bibitem [{\citenamefont {Schumayer}\ and\ \citenamefont {Hutchinson}(2011)}]{schumayer2011colloquium}%
  \BibitemOpen
  \bibfield  {author} {\bibinfo {author} {\bibfnamefont {D.}~\bibnamefont {Schumayer}}\ and\ \bibinfo {author} {\bibfnamefont {D.~A.}\ \bibnamefont {Hutchinson}},\ }\bibfield  {title} {\bibinfo {title} {Colloquium: Physics of the riemann hypothesis},\ }\href@noop {} {\bibfield  {journal} {\bibinfo  {journal} {Reviews of Modern Physics}\ }\textbf {\bibinfo {volume} {83}},\ \bibinfo {pages} {307} (\bibinfo {year} {2011})}\BibitemShut {NoStop}%
\bibitem [{\citenamefont {Wolf}(2020)}]{wolf2020will}%
  \BibitemOpen
  \bibfield  {author} {\bibinfo {author} {\bibfnamefont {M.}~\bibnamefont {Wolf}},\ }\bibfield  {title} {\bibinfo {title} {Will a physicist prove the riemann hypothesis?},\ }\href@noop {} {\bibfield  {journal} {\bibinfo  {journal} {Reports on Progress in Physics}\ }\textbf {\bibinfo {volume} {83}},\ \bibinfo {pages} {036001} (\bibinfo {year} {2020})}\BibitemShut {NoStop}%
\bibitem [{\citenamefont {He}\ \emph {et~al.}(2020)\citenamefont {He}, \citenamefont {Ai}, \citenamefont {Cui}, \citenamefont {Huang}, \citenamefont {Han}, \citenamefont {Li}, \citenamefont {Tu}, \citenamefont {Creffield}, \citenamefont {Sierra},\ and\ \citenamefont {Guo}}]{he2020identifying}%
  \BibitemOpen
  \bibfield  {author} {\bibinfo {author} {\bibfnamefont {R.}~\bibnamefont {He}}, \bibinfo {author} {\bibfnamefont {M.-Z.}\ \bibnamefont {Ai}}, \bibinfo {author} {\bibfnamefont {J.-M.}\ \bibnamefont {Cui}}, \bibinfo {author} {\bibfnamefont {Y.-F.}\ \bibnamefont {Huang}}, \bibinfo {author} {\bibfnamefont {Y.-J.}\ \bibnamefont {Han}}, \bibinfo {author} {\bibfnamefont {C.-F.}\ \bibnamefont {Li}}, \bibinfo {author} {\bibfnamefont {T.}~\bibnamefont {Tu}}, \bibinfo {author} {\bibfnamefont {C.}~\bibnamefont {Creffield}}, \bibinfo {author} {\bibfnamefont {G.}~\bibnamefont {Sierra}},\ and\ \bibinfo {author} {\bibfnamefont {G.-C.}\ \bibnamefont {Guo}},\ }\bibfield  {title} {\bibinfo {title} {Identifying the riemann zeros by periodically driving a single qubit},\ }\href@noop {} {\bibfield  {journal} {\bibinfo  {journal} {Physical Review A}\ }\textbf {\bibinfo {volume} {101}},\ \bibinfo {pages} {043402} (\bibinfo {year} {2020})}\BibitemShut {NoStop}%
\bibitem [{\citenamefont {Southier}\ \emph {et~al.}(2023)\citenamefont {Southier}, \citenamefont {Santos}, \citenamefont {Ribeiro},\ and\ \citenamefont {Ribeiro}}]{southier2023identifying}%
  \BibitemOpen
  \bibfield  {author} {\bibinfo {author} {\bibfnamefont {A.}~\bibnamefont {Southier}}, \bibinfo {author} {\bibfnamefont {L.~F.}\ \bibnamefont {Santos}}, \bibinfo {author} {\bibfnamefont {P.~S.}\ \bibnamefont {Ribeiro}},\ and\ \bibinfo {author} {\bibfnamefont {A.}~\bibnamefont {Ribeiro}},\ }\bibfield  {title} {\bibinfo {title} {Identifying primes from entanglement dynamics},\ }\href@noop {} {\bibfield  {journal} {\bibinfo  {journal} {Physical Review A}\ }\textbf {\bibinfo {volume} {108}},\ \bibinfo {pages} {042404} (\bibinfo {year} {2023})}\BibitemShut {NoStop}%
\bibitem [{\citenamefont {McGuigan}(2023)}]{mcguigan2023quantum}%
  \BibitemOpen
  \bibfield  {author} {\bibinfo {author} {\bibfnamefont {M.}~\bibnamefont {McGuigan}},\ }\bibfield  {title} {\bibinfo {title} {Quantum computing and the riemann hypothesis},\ }\href@noop {} {\bibfield  {journal} {\bibinfo  {journal} {arXiv preprint arXiv:2303.04602}\ } (\bibinfo {year} {2023})}\BibitemShut {NoStop}%
\bibitem [{\citenamefont {Berry}(1981)}]{berry1981quantizing}%
  \BibitemOpen
  \bibfield  {author} {\bibinfo {author} {\bibfnamefont {M.~V.}\ \bibnamefont {Berry}},\ }\bibfield  {title} {\bibinfo {title} {Quantizing a classically ergodic system: Sinai's billiard and the kkr method},\ }\href@noop {} {\bibfield  {journal} {\bibinfo  {journal} {Annals of Physics}\ }\textbf {\bibinfo {volume} {131}},\ \bibinfo {pages} {163} (\bibinfo {year} {1981})}\BibitemShut {NoStop}%
\bibitem [{\citenamefont {Bhaduri}\ \emph {et~al.}(1995)\citenamefont {Bhaduri}, \citenamefont {Khare},\ and\ \citenamefont {Law}}]{bhaduri1995phase}%
  \BibitemOpen
  \bibfield  {author} {\bibinfo {author} {\bibfnamefont {R.}~\bibnamefont {Bhaduri}}, \bibinfo {author} {\bibfnamefont {A.}~\bibnamefont {Khare}},\ and\ \bibinfo {author} {\bibfnamefont {J.}~\bibnamefont {Law}},\ }\bibfield  {title} {\bibinfo {title} {Phase of the riemann $\zeta$ function and the inverted harmonic oscillator},\ }\href@noop {} {\bibfield  {journal} {\bibinfo  {journal} {Physical Review E}\ }\textbf {\bibinfo {volume} {52}},\ \bibinfo {pages} {486} (\bibinfo {year} {1995})}\BibitemShut {NoStop}%
\bibitem [{\citenamefont {Moroz}(2010)}]{moroz2010below}%
  \BibitemOpen
  \bibfield  {author} {\bibinfo {author} {\bibfnamefont {S.}~\bibnamefont {Moroz}},\ }\bibfield  {title} {\bibinfo {title} {Below the breitenlohner-freedman bound in the nonrelativistic ads/cft correspondence},\ }\href@noop {} {\bibfield  {journal} {\bibinfo  {journal} {Physical Review D—Particles, Fields, Gravitation, and Cosmology}\ }\textbf {\bibinfo {volume} {81}},\ \bibinfo {pages} {066002} (\bibinfo {year} {2010})}\BibitemShut {NoStop}%
\bibitem [{\citenamefont {Berry}\ and\ \citenamefont {Keating}(1999{\natexlab{a}})}]{berry1999h}%
  \BibitemOpen
  \bibfield  {author} {\bibinfo {author} {\bibfnamefont {M.~V.}\ \bibnamefont {Berry}}\ and\ \bibinfo {author} {\bibfnamefont {J.~P.}\ \bibnamefont {Keating}},\ }\bibfield  {title} {\bibinfo {title} {H= xp and the riemann zeros},\ }in\ \href@noop {} {\emph {\bibinfo {booktitle} {Supersymmetry and trace formulae: chaos and disorder}}}\ (\bibinfo  {publisher} {Springer},\ \bibinfo {year} {1999})\ pp.\ \bibinfo {pages} {355--367}\BibitemShut {NoStop}%
\bibitem [{\citenamefont {Berry}\ and\ \citenamefont {Keating}(1999{\natexlab{b}})}]{berry1999riemann}%
  \BibitemOpen
  \bibfield  {author} {\bibinfo {author} {\bibfnamefont {M.~V.}\ \bibnamefont {Berry}}\ and\ \bibinfo {author} {\bibfnamefont {J.~P.}\ \bibnamefont {Keating}},\ }\bibfield  {title} {\bibinfo {title} {The riemann zeros and eigenvalue asymptotics},\ }\href@noop {} {\bibfield  {journal} {\bibinfo  {journal} {SIAM review}\ }\textbf {\bibinfo {volume} {41}},\ \bibinfo {pages} {236} (\bibinfo {year} {1999}{\natexlab{b}})}\BibitemShut {NoStop}%
\bibitem [{\citenamefont {Sundaram}\ \emph {et~al.}(2024)\citenamefont {Sundaram}, \citenamefont {Burgess},\ and\ \citenamefont {O’Dell}}]{sundaram2024duality}%
  \BibitemOpen
  \bibfield  {author} {\bibinfo {author} {\bibfnamefont {S.}~\bibnamefont {Sundaram}}, \bibinfo {author} {\bibfnamefont {C.}~\bibnamefont {Burgess}},\ and\ \bibinfo {author} {\bibfnamefont {D.}~\bibnamefont {O’Dell}},\ }\bibfield  {title} {\bibinfo {title} {Duality between the quantum inverted harmonic oscillator and inverse square potentials},\ }\href@noop {} {\bibfield  {journal} {\bibinfo  {journal} {New Journal of Physics}\ }\textbf {\bibinfo {volume} {26}},\ \bibinfo {pages} {053023} (\bibinfo {year} {2024})}\BibitemShut {NoStop}%
\bibitem [{\citenamefont {Birman}\ and\ \citenamefont {Krein}(1962)}]{birman1962theory}%
  \BibitemOpen
  \bibfield  {author} {\bibinfo {author} {\bibfnamefont {M.~S.}\ \bibnamefont {Birman}}\ and\ \bibinfo {author} {\bibfnamefont {M.}~\bibnamefont {Krein}},\ }\bibfield  {title} {\bibinfo {title} {On the theory of wave operators and scattering operators},\ }\href@noop {} {\bibfield  {journal} {\bibinfo  {journal} {Russian Mathematicians in the 20th Century}\ ,\ \bibinfo {pages} {463}} (\bibinfo {year} {1962})}\BibitemShut {NoStop}%
\bibitem [{\citenamefont {Wang}\ \emph {et~al.}(2014)\citenamefont {Wang}, \citenamefont {Stocks},\ and\ \citenamefont {Faulkner}}]{wang2014single}%
  \BibitemOpen
  \bibfield  {author} {\bibinfo {author} {\bibfnamefont {Y.}~\bibnamefont {Wang}}, \bibinfo {author} {\bibfnamefont {G.~M.}\ \bibnamefont {Stocks}},\ and\ \bibinfo {author} {\bibfnamefont {J.~S.}\ \bibnamefont {Faulkner}},\ }\bibfield  {title} {\bibinfo {title} {The single-site green’s function and krein’s theorem},\ }\href@noop {} {\bibfield  {journal} {\bibinfo  {journal} {Journal of Physics: Condensed Matter}\ }\textbf {\bibinfo {volume} {26}},\ \bibinfo {pages} {274208} (\bibinfo {year} {2014})}\BibitemShut {NoStop}%
\bibitem [{\citenamefont {Galkowski}\ \emph {et~al.}(2024)\citenamefont {Galkowski}, \citenamefont {Marchand}, \citenamefont {Wang},\ and\ \citenamefont {Zworski}}]{Galkowski2024}%
  \BibitemOpen
  \bibfield  {author} {\bibinfo {author} {\bibfnamefont {J.}~\bibnamefont {Galkowski}}, \bibinfo {author} {\bibfnamefont {P.}~\bibnamefont {Marchand}}, \bibinfo {author} {\bibfnamefont {J.}~\bibnamefont {Wang}},\ and\ \bibinfo {author} {\bibfnamefont {M.}~\bibnamefont {Zworski}},\ }\bibfield  {title} {\bibinfo {title} {The scattering phase: Seen at last},\ }\href {https://doi.org/10.1137/23M1547147} {\bibfield  {journal} {\bibinfo  {journal} {SIAM Journal on Applied Mathematics}\ }\textbf {\bibinfo {volume} {84}},\ \bibinfo {pages} {246} (\bibinfo {year} {2024})},\ \Eprint {https://arxiv.org/abs/https://doi.org/10.1137/23M1547147} {https://doi.org/10.1137/23M1547147} \BibitemShut {NoStop}%
\bibitem [{\citenamefont {Faulkner}(1977)}]{faulkner1977scattering}%
  \BibitemOpen
  \bibfield  {author} {\bibinfo {author} {\bibfnamefont {J.}~\bibnamefont {Faulkner}},\ }\bibfield  {title} {\bibinfo {title} {Scattering theory and cluster calculations},\ }\href@noop {} {\bibfield  {journal} {\bibinfo  {journal} {Journal of Physics C: Solid State Physics}\ }\textbf {\bibinfo {volume} {10}},\ \bibinfo {pages} {4661} (\bibinfo {year} {1977})}\BibitemShut {NoStop}%
\bibitem [{\citenamefont {Gupta}\ \emph {et~al.}(2013)\citenamefont {Gupta}, \citenamefont {Harikumar},\ and\ \citenamefont {de~Queiroz}}]{gupta2013dirac}%
  \BibitemOpen
  \bibfield  {author} {\bibinfo {author} {\bibfnamefont {K.~S.}\ \bibnamefont {Gupta}}, \bibinfo {author} {\bibfnamefont {E.}~\bibnamefont {Harikumar}},\ and\ \bibinfo {author} {\bibfnamefont {A.~R.}\ \bibnamefont {de~Queiroz}},\ }\bibfield  {title} {\bibinfo {title} {A dirac-type variant of the xp-model and the riemann zeros},\ }\href@noop {} {\bibfield  {journal} {\bibinfo  {journal} {Europhysics Letters}\ }\textbf {\bibinfo {volume} {102}},\ \bibinfo {pages} {10006} (\bibinfo {year} {2013})}\BibitemShut {NoStop}%
\bibitem [{\citenamefont {Rodr{\'\i}guez-Laguna}\ \emph {et~al.}(2017)\citenamefont {Rodr{\'\i}guez-Laguna}, \citenamefont {Tarruell}, \citenamefont {Lewenstein},\ and\ \citenamefont {Celi}}]{rodriguez2017synthetic}%
  \BibitemOpen
  \bibfield  {author} {\bibinfo {author} {\bibfnamefont {J.}~\bibnamefont {Rodr{\'\i}guez-Laguna}}, \bibinfo {author} {\bibfnamefont {L.}~\bibnamefont {Tarruell}}, \bibinfo {author} {\bibfnamefont {M.}~\bibnamefont {Lewenstein}},\ and\ \bibinfo {author} {\bibfnamefont {A.}~\bibnamefont {Celi}},\ }\bibfield  {title} {\bibinfo {title} {Synthetic unruh effect in cold atoms},\ }\href@noop {} {\bibfield  {journal} {\bibinfo  {journal} {Physical Review A}\ }\textbf {\bibinfo {volume} {95}},\ \bibinfo {pages} {013627} (\bibinfo {year} {2017})}\BibitemShut {NoStop}%
\bibitem [{\citenamefont {Elliott}\ and\ \citenamefont {Franz}(2015)}]{elliott2015colloquium}%
  \BibitemOpen
  \bibfield  {author} {\bibinfo {author} {\bibfnamefont {S.~R.}\ \bibnamefont {Elliott}}\ and\ \bibinfo {author} {\bibfnamefont {M.}~\bibnamefont {Franz}},\ }\bibfield  {title} {\bibinfo {title} {Colloquium: Majorana fermions in nuclear, particle, and solid-state physics},\ }\href@noop {} {\bibfield  {journal} {\bibinfo  {journal} {Reviews of Modern Physics}\ }\textbf {\bibinfo {volume} {87}},\ \bibinfo {pages} {137} (\bibinfo {year} {2015})}\BibitemShut {NoStop}%
\bibitem [{\citenamefont {Flensberg}\ \emph {et~al.}(2021)\citenamefont {Flensberg}, \citenamefont {von Oppen},\ and\ \citenamefont {Stern}}]{flensberg2021engineered}%
  \BibitemOpen
  \bibfield  {author} {\bibinfo {author} {\bibfnamefont {K.}~\bibnamefont {Flensberg}}, \bibinfo {author} {\bibfnamefont {F.}~\bibnamefont {von Oppen}},\ and\ \bibinfo {author} {\bibfnamefont {A.}~\bibnamefont {Stern}},\ }\bibfield  {title} {\bibinfo {title} {Engineered platforms for topological superconductivity and majorana zero modes},\ }\href@noop {} {\bibfield  {journal} {\bibinfo  {journal} {Nature Reviews Materials}\ }\textbf {\bibinfo {volume} {6}},\ \bibinfo {pages} {944} (\bibinfo {year} {2021})}\BibitemShut {NoStop}%
\bibitem [{\citenamefont {Berry}(1984)}]{berry1984quantal}%
  \BibitemOpen
  \bibfield  {author} {\bibinfo {author} {\bibfnamefont {M.~V.}\ \bibnamefont {Berry}},\ }\bibfield  {title} {\bibinfo {title} {Quantal phase factors accompanying adiabatic changes},\ }\href@noop {} {\bibfield  {journal} {\bibinfo  {journal} {Proceedings of the Royal Society of London. A. Mathematical and Physical Sciences}\ }\textbf {\bibinfo {volume} {392}},\ \bibinfo {pages} {45} (\bibinfo {year} {1984})}\BibitemShut {NoStop}%
\end{thebibliography}

\end{document}